\documentclass[preprint,nofootinbib,preprintnumbers,amsmath,amssymb]{revtex4}
\usepackage{graphicx}% Include figure files
\usepackage{dcolumn}% Align table columns on decimal point
\usepackage{bm}% bold math

\RequirePackage{xspace}
\usepackage{relsize}

\def\Kbar  {\kern 0.2em\overline{\kern -0.2em K}{}\xspace}

\def\Bbar    {\kern 0.18em\overline{\kern -0.18em B}{}\xspace}

\def\missET {{\not\!\! E_t}}
\def\Qbar    {\kern 0.08em\overline{\kern -0.08em Q}{}\xspace}

\newcommand{\mev}{\ensuremath{\mathrm{\,Me\kern -0.1em V}}\xspace}
\newcommand{\mevc}{\ensuremath{{\mathrm{\,Me\kern -0.1em V\!/}c}}\xspace}
\newcommand{\mevcc}{\ensuremath{{\mathrm{\,Me\kern -0.1em V\!/}c^2}}\xspace}
\newcommand{\gev}{\ensuremath{\mathrm{\,Ge\kern -0.1em V}}\xspace}
\newcommand{\gevc}{\ensuremath{{\mathrm{\,Ge\kern -0.1em V\!/}c}}\xspace}
\newcommand{\gevcnospace}{\ensuremath{{\mathrm{\,Ge\kern -0.1em V\!/}c}}}
\newcommand{\gevcc}{\ensuremath{{\mathrm{\,Ge\kern -0.1em V\!/}c^2}}\xspace}
\newcommand{\be}{\begin{equation}}
\newcommand{\ee}{\end{equation}}
\newcommand{\benn}{\begin{equation*}}
\newcommand{\eenn}{\end{equation*}}
\newcommand{\bea}{\begin{eqnarray}}
\newcommand{\eea}{\end{eqnarray}}

\def\beq{\begin{equation}}
\def\eeq{\end{equation}}
\def\bea{\begin{eqnarray}}
\def\eea{\end{eqnarray}}

\newcommand{\g}{\gamma}

\usepackage{color}

\def\missET {{\not\!\! E_T}}

\begin{document}

\preprint{UCI-HEP-TR-2010-15}

\title{ Constraints on Dark Matter from Colliders}
%\title{{Search for fermion-pair decays}}
%\\ \vspace*{2.0cm}}

\author{Jessica~Goodman, Masahiro~Ibe, Arvind~Rajaraman,
William~Shepherd, Tim M.P.~Tait and Hai-Bo Yu
\\
{\small {\it Department of Physics and Astronomy,}}
\\
{\small {\it University of California, Irvine, California 92697}}}
%\affiliation{Department of Physics \& Astronomy, University of California, Irvine, CA 92697}
\date{\today}

\begin{abstract}
We show that colliders can impose strong constraints
on models of dark matter, in particular when the dark matter is light.
We analyze models where the dark matter is a fermion or scalar interacting
with quarks and/or gluons through an effective theory containing
higher dimensional operators which represent heavier states that have
been integrated out of the effective field theory.
We determine bounds from existing Tevatron searches for monojets as well as
expected LHC reaches for a discovery.  We find that colliders can provide
information which is complementary or in some cases even superior to
experiments searching for direct detection of dark matter through its scattering
with nuclei. In particular, both the Tevatron and the LHC can
outperform spin dependent searches by an
order of magnitude or better over much of parameter space, and if the dark matter
couples mainly to gluons, the LHC can place bounds  superior to any spin independent search.
\end{abstract}

\maketitle
\section{Introduction}
\label{sec:intro}

While astrophysical observations provide compelling proof
for the existence of a non-baryonic dark component to the Universe and precise
measurements as to its abundance \cite{Komatsu:2010fb},
they offer no clue as to the mass of dark matter (DM) particles, how they fit into the Standard Model
(SM) of particle physics, or even whether or not the dark matter has interactions
beyond gravitational.
The most compelling vision of dark matter is a weakly interacting massive particle (WIMP),
which offers the possibility to understand the relic abundance of dark matter
as a natural consequence of the thermal history of the Universe through the WIMP(less) miracle
\cite{Feng:2008ya}.
The large
interactions of WIMPs with SM particles may imply detectable rates
of WIMP annihilations into SM final states, scattering of WIMPs with heavy nuclei, and
production of WIMPs in high energy reactions of SM particles at colliders.

WIMPs which produce signals in direct detection experiments must also couple to nucleons,
and therefore can be produced at colliders like the Tevatron and LHC.
Low mass particles are particularly
amenable to searches at colliders, since the fact that a typical collision involves quarks and/or
gluons carrying only a small fraction of the parent (anti-)proton energy implies that cross
sections fall dramatically with the mass of produced states.
Light states can thus be produced
with very large rates.  In the case of a WIMP, stability on the order of the lifetime of the Universe
implies that pair production must highly dominate over single production, and precludes the WIMP
from decaying within the detector volume.
WIMPs therefore appear as missing energy, and can potentially
be observed by searching for visible particles recoiling against dark matter particles
\cite{Birkedal:2004xn,Beltran:2008xg,Shepherd:2009sa,Cao:2009uw,Beltran:2010ww}.
This can be used to set constraints on the WIMP couplings to the constituents of nuclei, which in turn
can be translated to constraints on direct detection cross sections. In previous work~\cite{Goodman:2010yf},
this was done for the case of Majorana
WIMPs\footnote{A partial set of operators for a Dirac WIMP were bounded using Tevatron
data in Ref.~\cite{Bai:2010hh} and the Tevatron limits and LHC reach for operator
D8 (see below) were considered in Ref.~\cite{Beltran:2010ww}.};
 here we extend this work to also include Dirac fermion and scalar (real or complex) WIMPs.

There is currently particular interest in light dark matter particles.
The DAMA experiment has reported a signal of annual modulation
at a high significance level \cite{Bernabei:2010mq}; this signal is consistent with a dark matter
discovery
interpretation with a dark matter particle of mass $\lesssim 10$~GeV
\cite{Petriello:2008jj,Feng:2008dz}. The CoGeNT collaboration has also reported a signal \cite{Aalseth:2010vx}
which can be explained by a WIMP in this mass range (though there may be some tension
with unpublished data from 5 towers of CDMS Si detectors \cite{CDMS-Si}, and
with recent data from XENON10/100 \cite{Aprile:2010um,Sorensen:2010hq}).
There has hence been much recent interest in models of light dark matter (where the DM mass is
order a few GeV)
\cite{Kim:2009ke,Fitzpatrick:2010em,Kopp:2009qt,Kuflik:2010ah,Chang:2010yk,
Essig:2010ye,An:2010kc,Andreas:2010dz,Barger:2010yn,Hooper:2010uy}.
As colliders are most effective when producing highly boosted, light WIMPs, the
tantalizing hints from DAMA and CoGeNT point toward a region where colliders can have
a particular impact on theories of dark matter.

In this article, we consider the situation
where the WIMP is the only new particle in the energy ranges
relevant for current experiments. Given the small energy
transfers involved in direct detection, this assumption is almost certainly justified.
For colliders,
the degree to which it is justified depends on the details of the WIMP theory.   Under this assumption, the WIMP will couple to the SM particles through higher dimensional operators,
presumably mediated by particles of the dark sector which are somewhat heavier than the
WIMP itself (and which may or may not carry SM gauge charges).

This article is organized as follows.  In Section~\ref{sec:couplings},
we write down a complete list of leading operators and analyze
the constraints on the coefficients of these operators, assuming that only one operator
is dominant at a time. In Section~\ref{sec:collider}, we employ existing Tevatron and planned LHC searches
to determine the constraints on the coefficients of the operators (or prospects for their discovery),
respectively.  Of the complete set of operators, some mediate substantial (i.e. not suppressed
by the WIMP velocity) low energy WIMP-nucleus rates, and thus are constrained
or may be discovered by direct detection experiments.  These bounds and prospects are
presented in Section~\ref{sec:direct}.
We conclude with comments on future directions in Section~\ref{sec:outlook}.

 \section{Effective Field Theory of WIMP couplings}
 \label{sec:couplings}

We consider the cases where the DM particle is a scalar or a fermion; if a scalar, it can be
real or complex, and if a fermion, it can be Majorana or Dirac.
Each of these cases is considered separately.
We note that in principle, the WIMP could also be
spin one or higher; we shall not consider these cases here since the couplings of such WIMPs
are usually restricted by gauge invariance and other symmetries, and are more heavily
model-dependent.

We shall be considering the situation where the WIMP (which we will generically
denote $\chi$) is the only
particle in addition to the standard model fields accessible to colliders.
We will assume that $\chi$ is odd under some $Z_2$ symmetry
(e.g. R-parity in supersymmetry, or KK-parity in extra dimensions),
and hence each coupling involves an even number of WIMPs with the lowest dimensional
operators we consider containing two WIMPs.
We assume whatever particles mediate interactions between the WIMPs and the SM fields
are somewhat heavier than the WIMPs themselves, with their leading effect manifest
 as higher dimensional operators in the effective field theory.
For simplicity, we assume the WIMP is a singlet under the SM gauge groups, and thus
possesses no tree-level couplings to the electroweak gauge bosons.  We also
neglect couplings with Higgs bosons.
While the inclusion of
such couplings in the effective theory is straightforward, we leave these cases for future work.
Given the assumption that the WIMPs are SM singlets, the factor in each operator
consisting of SM fields must also be invariant under SM gauge transformations.

We note in passing that even for an electroweak singlet WIMP, the lowest dimensional
operator linking a pair of WIMPs to the SM fields contains two WIMPs and the SM
Higgs bilinear $|H|^2$ \cite{Burgess:2000yq}.  Such an interaction
contributes to direct detection and collider processes involving WIMPs
by inducing a $\chi$-$\chi$-$h^0$ interaction after electroweak
symmetry-breaking\footnote{Collider and direct detection signals
from this operator are explored in Ref.~\cite{Kanemura:2010sh}.}.
While we do not consider this operator further, we note that
for cases where the Higgs is heavy enough, it is effectively integrated out, leaving
behind operators which we do consider.

The next allowed class of operators have SM factors which are
quark or lepton bilinears.  The lepton bilinear couplings contribute only at a very suppressed
level to direct detection or hadron collider production, leaving us with little to say about them.
It would be very interesting to study constraints on such operators from
indirect detection experiments and/or lepton colliders such as LEP-II. In this note
we shall focus on quark bilinear operators of the form $\bar{q}\Gamma q$,
where $\Gamma$ is a $4 \times 4$ matrix of the complete set,
\bea
\Gamma & = &
\left\{ 1, \gamma^5, \gamma^\mu, \gamma^\mu \gamma^5, \sigma^{\mu \nu} \right\} ~.
\eea
We do not consider terms with derivatives acting on the quarks, which lead to higher
dimensional operators, more suppressed at low energies.

Finally, we have operators mediating WIMP couplings to massless
gauge fields. The leading operators are
a magnetic moment coupling $\bar{\chi}\sigma^{\mu\nu}\chi F_{\mu\nu}$ and an electric dipole
moment coupling $\bar{\chi}\sigma^{\mu\nu}\gamma_5\chi F_{\mu\nu}$
(which are only non-vanishing for a Dirac fermion WIMP), though given the unbroken
$U(1)_{EM}$ gauge invariance, they are likely induced at the loop level and thus may
have small coefficients.
Various experimental bounds and direct detection signals of these operators have been studied in
Refs.~\cite{Fitzpatrick:2010br,Bagnasco:1993st,Pospelov:2000bq,
Sigurdson:2004zp,Gardner:2008yn,Masso:2009mu,Cho:2010br,Chang:2010en,Barger:2010gv,Banks:2010eh}.
We do not consider collider constraints for these operators further here and leave astrophysics
bounds for future work~\cite{linesearch}. We also have couplings to  $G_{\mu\nu}G_{\alpha\beta}
$, where $G G$ can either be a pair of electromagnetic or color field strengths, with
gauge and Lorentz indices contracted in all possible ways to form a family of related operators.
Here we focus on the operators involving color field strengths.
Just as for quark operators, terms with derivatives acting on the gauge field strengths are higher
order and more suppressed.

All together, these higher-dimensional operators define an effective field theory of the
interactions of singlet WIMPs with hadronic matter.
It is expected to reasonably capture
the physics provided the WIMP is somewhat lighter than the particles which mediate its
interactions with the SM.
It is a non-renormalizable field theory and thus must break down at some energy scale,
represented by the masses of those particles which have been integrated out.
The quantities $M_*$ which characterize the interaction strength of the interactions
are functions of the masses and the coupling strengths of
the mediating particles to WIMPs and SM fields, and can be computed in terms of the
fundamental parameters for any specific UV theory of interest.

What happens above the regime of validity of the effective theory
depends on the UV completion, and is much more model dependent.
Depending on the specifics of the UV theory, collider bounds may get stronger
or weaker.  For example, in
supersymmetric theories our operators are UV completed into
colored squarks which will be produced on-shell and may contribute to the jets + missing
energy observable with {\em larger} rates than those we are computing in the effective theory.
Other UV completions, such as a light neutral mediator, can lead to much {\em weaker} collider
cross sections \cite{Bai:2010hh}, since far above the mediator mass the rate will fall with jet
transverse energy as $1 / P^2_t$, whereas in the effective theory the partonic reaction is
flat with jet $P_t$, scaling as $1 / M_*^2$.  Thus, it should be borne in mind that our limits
strictly speaking only apply when all mediator masses are much larger than the typical
energy of the reaction, and in the absence of a picture of the UV theory, it is hard to know whether
the bounds are over- or under-estimated when the effective theory description does not strictly
apply.

For a given WIMP mass, there is a lower bound on $M_*$ such that one can imagine any
weakly coupled UV completion.  Since the operators mediate interactions with (at least)
two colored SM fields coupled to two WIMPs, the simplest tree level UV completions have a single
mediator particle and two interactions.  The mapping to $M_*$ from the UV parameters thus
involves an expression such as $M_* \sim M /\sqrt{g_1 g_2}$ where $M$ is the mass of the exchanged particle, and $g_1$ and $g_2$ are couplings.  Since an effective theory
description requires $M > 2m_\chi$, and a preturbative
theory $g_1 g_2 \lesssim (4 \pi)^{2}$, a weakly coupled
UV completion requires $m_\chi \lesssim 2\pi M_*$, beyond which the
UV completion becomes
non-perturbative.  In determining bounds, since there is no imaginable
perturbative UV picture for $m_\chi \lesssim 2\pi M_*$, we cut off the bounded regions
outside of this  region of validity. Furthermore, for the effective theory to make sense, the mediator mass has to be larger than energy transfer through quarks at the collider environment. The limit, in which the effective theory breaks down, highly depends on the details of relevant patron energy and its distribution. Since $M\lesssim 4\pi M_*$ for the perturbative UV completion, our bounds are valid when the characteristic energy transfer is smaller than $4\pi M_*$. The detailed analysis of this limit is beyond the scope of this work, we will leave it for the future investigation.

\begin{table}
 \hspace{0.033\textwidth}
 \begin{minipage}{0.4\textwidth}
  \centering
 \begin{tabular}{|c|c|c|}
\hline
   Name    & Operator & Coefficient  \\
\hline
D1 & $\bar{\chi}\chi\bar{q} q$ & $m_q/M_*^3$   \\
D2 & $\bar{\chi}\g^5\chi\bar{q} q$ & $im_q/M_*^3$    \\
D3 & $\bar{\chi}\chi\bar{q}\g^5 q$ & $im_q/M_*^3$    \\
D4 & $\bar{\chi}\g^5\chi\bar{q}\g^5 q$ & $m_q/M_*^3$   \\
D5 & $\bar{\chi}\g^{\mu}\chi\bar{q}\g_{\mu} q$ & $1/M_*^2$   \\
D6 & $\bar{\chi}\g^{\mu}\g^5\chi\bar{q}\g_{\mu} q$ & $1/M_*^2$    \\
D7 & $\bar{\chi}\g^{\mu}\chi\bar{q}\g_{\mu}\g^5 q$ & $1/M_*^2$   \\
D8 & $\bar{\chi}\g^{\mu}\g^5\chi\bar{q}\g_{\mu}\g^5 q$ & $1/M_*^2$   \\
D9 & $\bar{\chi}\sigma^{\mu\nu}\chi\bar{q}\sigma_{\mu\nu} q$ & $1/M_*^2$   \\
D10 & $\bar{\chi}\sigma_{\mu\nu}\g^5\chi\bar{q}\sigma_{\alpha\beta}q$ & $i/M_*^2$  \\
D11 & $\bar{\chi}\chi G_{\mu\nu}G^{\mu\nu}$ & $\alpha_s/4M_*^3$   \\
D12 & $\bar{\chi}\g^5\chi G_{\mu\nu}G^{\mu\nu}$ & $i\alpha_s/4M_*^3$   \\
D13 & $\bar{\chi}\chi G_{\mu\nu}\tilde{G}^{\mu\nu}$ & $i\alpha_s/4M_*^3$   \\
D14 & $\bar{\chi}\g^5\chi G_{\mu\nu}\tilde{G}^{\mu\nu}$  & $\alpha_s/4M_*^3$ \\
\hline
\end{tabular}
 % \caption{\textnormal{Caption 1}}
  \label{tab:caption1}
 \end{minipage}
 \hspace{0.033\textwidth}
 \begin{minipage}{0.4\textwidth}
  \centering
  \vspace*{-2.2cm}
\begin{tabular}{|c|c|c|}
\hline
   Name    & Operator & Coefficient   \\
\hline
C1 & $\chi^\dagger\chi\bar{q}q$ & $m_q/M_*^2$    \\
C2 & $\chi^\dagger\chi\bar{q}\g^5 q$ & $im_q/M_*^2$   \\
C3 &  $\chi^\dagger\partial_\mu\chi\bar{q}\g^\mu q$ & $1/M_*^2$   \\
C4 &  $\chi^\dagger\partial_\mu\chi\bar{q}\g^\mu\g^5q$ & $1/M_*^2$    \\
C5 & $\chi^\dagger\chi G_{\mu\nu}G^{\mu\nu}$  & $\alpha_s/4M_*^2$   \\
C6 & $\chi^\dagger\chi G_{\mu\nu}\tilde{G}^{\mu\nu}$  & $i\alpha_s/4M_*^2$   \\ \hline
R1 & $\chi^2\bar{q}q$ & $m_q/2M_*^2$    \\
R2 & $\chi^2\bar{q}\g^5 q$ & $im_q/2M_*^2$   \\
R3 & $\chi^2 G_{\mu\nu}G^{\mu\nu}$  & $\alpha_s/8M_*^2$   \\
R4 & $\chi^2 G_{\mu\nu}\tilde{G}^{\mu\nu}$  & $i\alpha_s/8M_*^2$   \\
\hline
\end{tabular}
  \label{tab:caption2}
 \end{minipage}
   \caption{\textnormal{Operators coupling WIMPs to SM particles. The operator names beginning with D, C, R  apply to WIMPS that are Dirac fermions, complex scalars or real scalars respectively. }}
\end{table}

The coefficients of the operators are chosen to simplify
comparisons to direct detection experiments. For quark bilinears, the appropriate
matrix elements (at low momentum transfer)
are $\langle N | m_q \bar{q} q | N \rangle$ and $\langle N | \bar{q} \gamma^\mu q | N \rangle$
which contribute to spin-independent scattering,
$\langle N | \bar{q} \gamma^\mu \gamma^5 q | N \rangle$, which contributes to spin-dependent
scattering, and $\langle N | \bar{q}\sigma^{\mu\nu} q | N \rangle$, which
couples to the magnetic moment of the nucleon.
For the gluon operators, the relevant matrix element is $\langle N | \alpha_s G G | N \rangle$.
The scalar (and pseudo-scalar) quark bilinears are normalized
by $m_q$, which together with our choice of universal
vector-type couplings has the added feature of mitigating contributions to
flavor changing processes from these operators, through the framework
of minimal flavor violation \cite{Buras:2000dm}. For the gluon field strength operators,
we normalize by a factor $\alpha_s$, which both anticipates their origin as loop processes
and captures the dominant renormalization group evolution.
The complete list of leading operators is given in Table I. The
coefficients of these operators have been scaled by appropriate powers
of $M_*$ (the value of which can be in principle different for each operator) to give the correct
over-all dimension in the action.

\section{Collider Constraints}
\label{sec:collider}

\subsection{Overview}

We can constrain $M_*$ for each operator in the table above by considering
 the pair production of  WIMPs at a hadron collider:
\begin{eqnarray}
 p\bar{p}\, (pp) \to \chi\chi + X%+ {\rm jets}
 .
\end{eqnarray}
Since the WIMPs escape undetected, this leads to events with missing transverse energy,
recoiling against additional hadronic radiation present in the reaction.

The most significant Standard Model backgrounds to this process are events where a
$Z$ boson decays into neutrinos,  together
with the associated production of jets. This background is irreducible.
There are also backgrounds  from events where a particle is either missed or
has a mismeasured energy. The most important of these comes from events producing
$W$ + jets, where the charged
lepton from the $W$-decay is missed. Other backgrounds such as QCD multijet production
(with the missing energy the result of mismeasuring the energy of one of more jets)
are expected to be subdominant for the cuts chosen in the analyses \cite{Alwall:2008va,Aaltonen:2008hh}.

\subsection{Tevatron Constraints}

The Tevatron has searched for signals of new physics with missing transverse energy
in many channels. We
will focus on monojet events,
where the WIMPs recoil against a single jet, with restrictions on any additional SM radiation.
We will compare the predictions of our effective theories
with the results on monojet events from CDF \,\cite{Aaltonen:2008hh,CDF}.
We expect similar constraints can be derived from D0 data, but choose to focus on the
available CDF searches which utilize much greater integrated luminosity.
In Ref.\,\cite{CDF}, the events were required to satisfy:
\begin{itemize}
\item  Events are required to have a leading Jet with transverse energy $E_t>80$\,GeV;
\item  Events must have net missing transverse energy $E_t>80$\,GeV;
\item  A second jet with $E_t<30$\,GeV is allowed;
\item  Events containing additional jets with $E_t>20$\,GeV are vetoed.
\end{itemize}
%and used the data samples corresponding to

In order to simulate WIMP pair production events to compare to these bounds,
we found the partonic cross section for $p\bar{p}\to j\chi\chi$ using
Comphep\,\cite{Pukhov:1999gg,Boos:2004kh}, where
$j$ is any parton other than the top quark, and is required to have
$E_t>80$\,GeV. At the parton level,
this simultaneously requires that $\missET>80$\,GeV. We correct
these parton-level estimates by an efficiency taking into account corrections from
parton showering, hadronization, and energy smearing by the detector.  This efficiency
is computed by first hadronizing the generated parton-level events
using Pythia\,\cite{Sjostrand:2006za} (through the Comphep-Pythia
interface\,\cite{Belyaev:2000wn}).  The hadronized events are reconstructed at the
detector level by passing them through PGS~\cite{PGS} tuned to simulate the response
of the CDF detector\footnote{
A previous study \cite{Beltran:2010ww} found that this detector model was able to reproduce the
backgrounds quoted in \cite{CDF} to the few percent level.} and
required to satisfy the detailed CDF analysis cuts.
The efficiency is defined as the ratio of the number of events after the PGS-level cuts
to the number at the parton-level.
We find that efficiencies range from $30\%$ to $48\%$ for %complex scalar and Dirac
WIMPs with various spins and masses ranging
from $0-300$ {\rm GeV}. Given the relative insensitivity to the details,
we choose a flat efficiency of $40\%$ for all cases.

\begin{figure}[h]
\includegraphics[width=12.0cm]{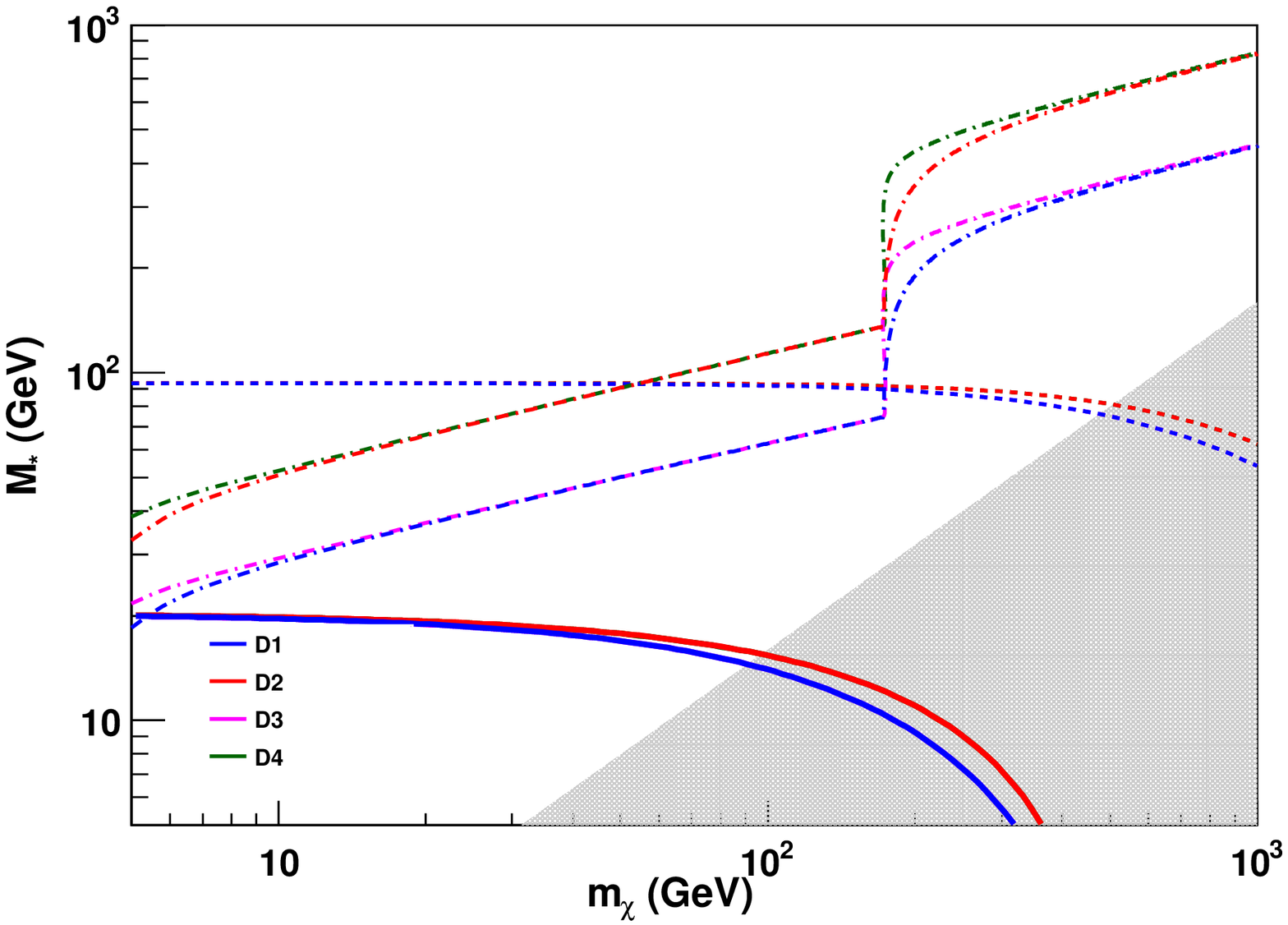}
\caption{\label{fig:D1-4}
Constraints on the operators D1-4. Solid lines are Tevatron $2\sigma$ lower limits, dashed lines
are LHC $5\sigma$ discovery reach lower limits, and dot-dashed lines indicate the values
of $M_*$ necessary for the WIMP to have the correct relic abundance in absence of any other
interactions. The curves for operators D1 and D2 are largely degenerate with those for D3 and D4,
respectively. The gray filled region indicates where the effective field theory breaks down,
possessing no simple perturbative UV completion.
}
\end{figure}

\begin{figure}[h]
\includegraphics[width=12.0cm]{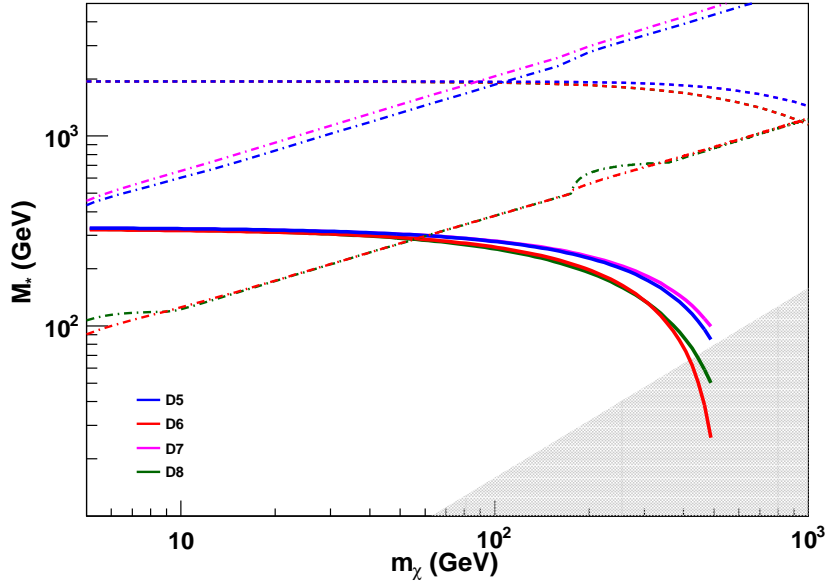}
\caption{\label{fig:D5-8}
Same as Fig.\,\ref{fig:D1-4}, but for the operators D5 and D6 which are largely degenerate with D7 and D8, respectively.
}
\end{figure}

\begin{figure}[h]
\includegraphics[width=12.0cm]{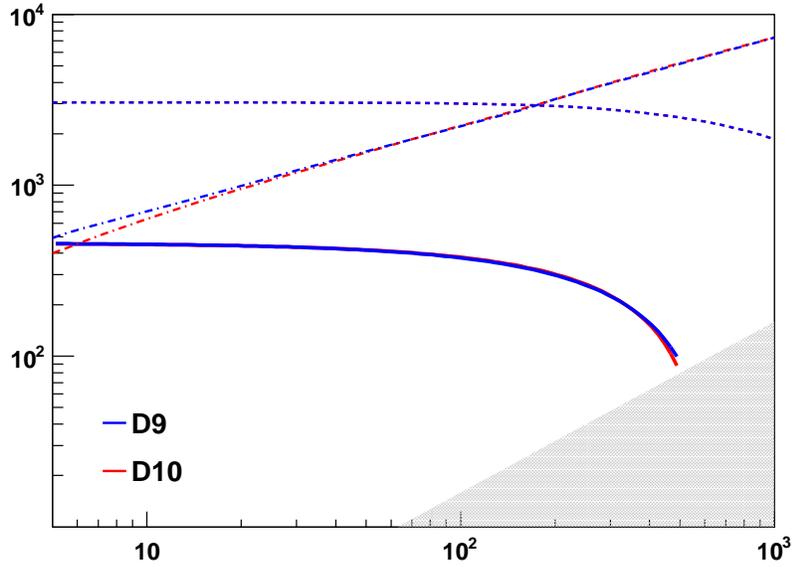}
\caption{\label{fig:D9-10}
Same as Fig.\,\ref{fig:D1-4}, but for the operators D9 and D10.
}
\end{figure}

\begin{figure}[h]
\includegraphics[width=12.0cm]{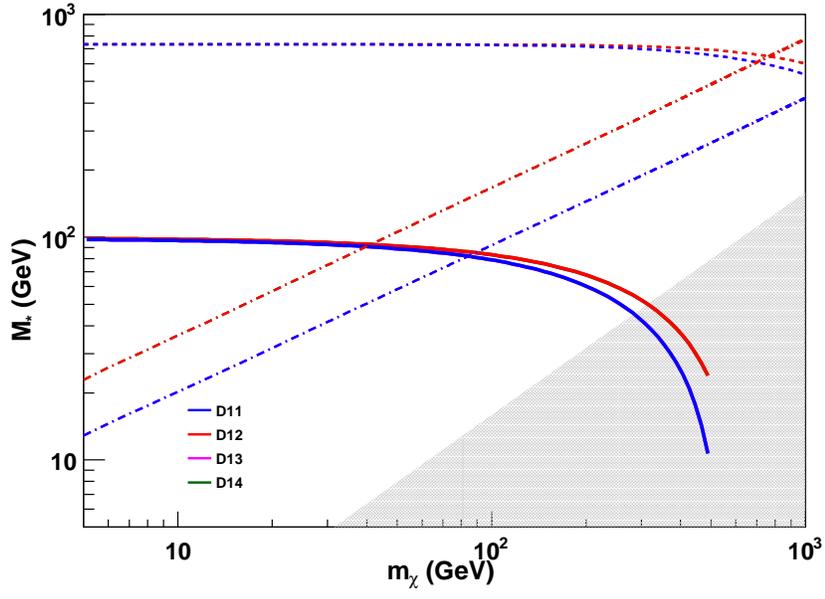}
\caption{\label{fig:D11-14}
Same as Fig.\,\ref{fig:D1-4}, but for the operators D11 and D12 which are largely degenerate with D13 and D14, respectively.
}
\end{figure}

\begin{figure}[h]
\includegraphics[width=12.0cm]{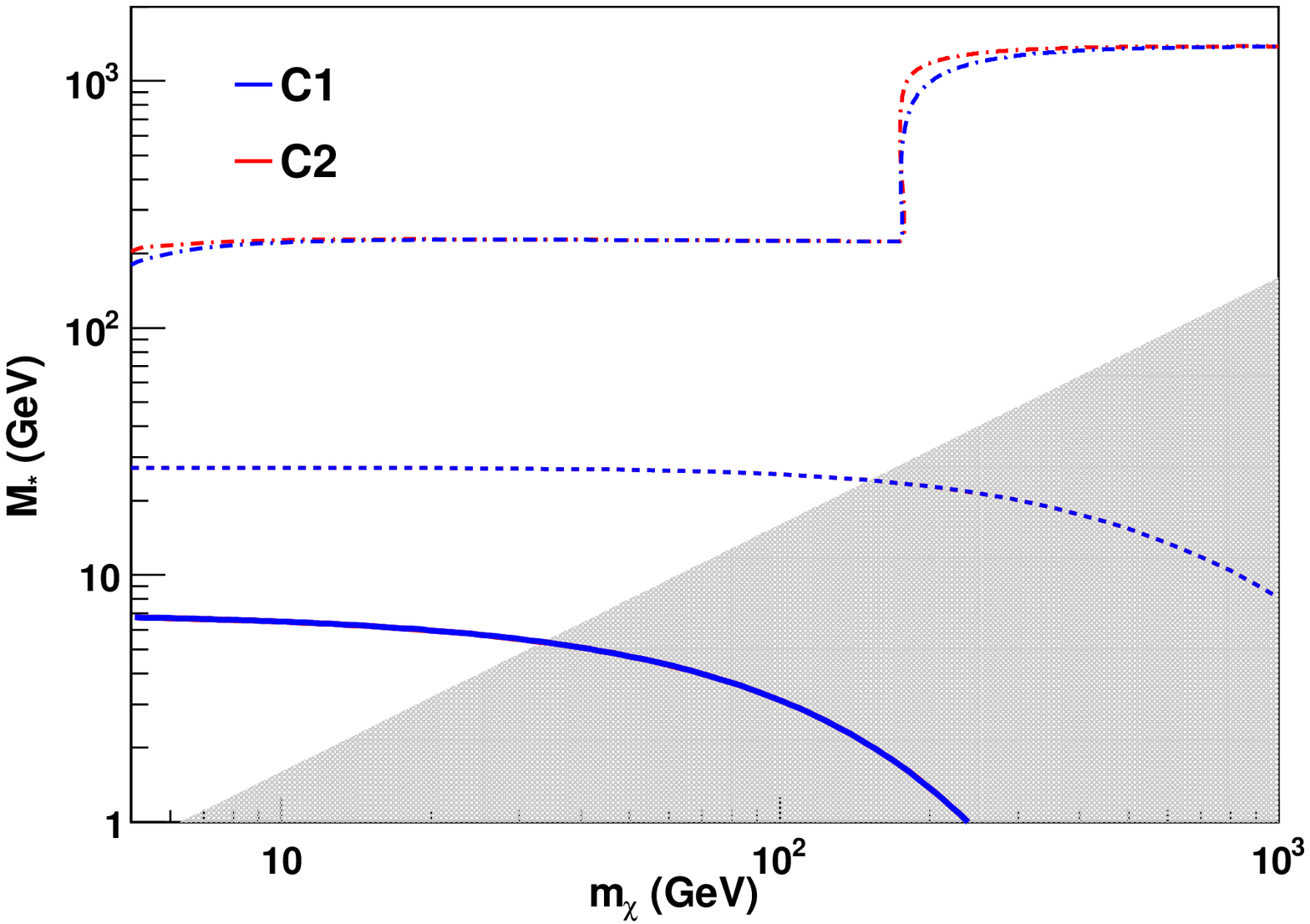}
\caption{\label{fig:C1-2}
Same as Fig.\,\ref{fig:D1-4}, but for the largely degenerate operators C1 and C2.
}
\end{figure}

\begin{figure}[t]
\includegraphics[width=12.0cm]{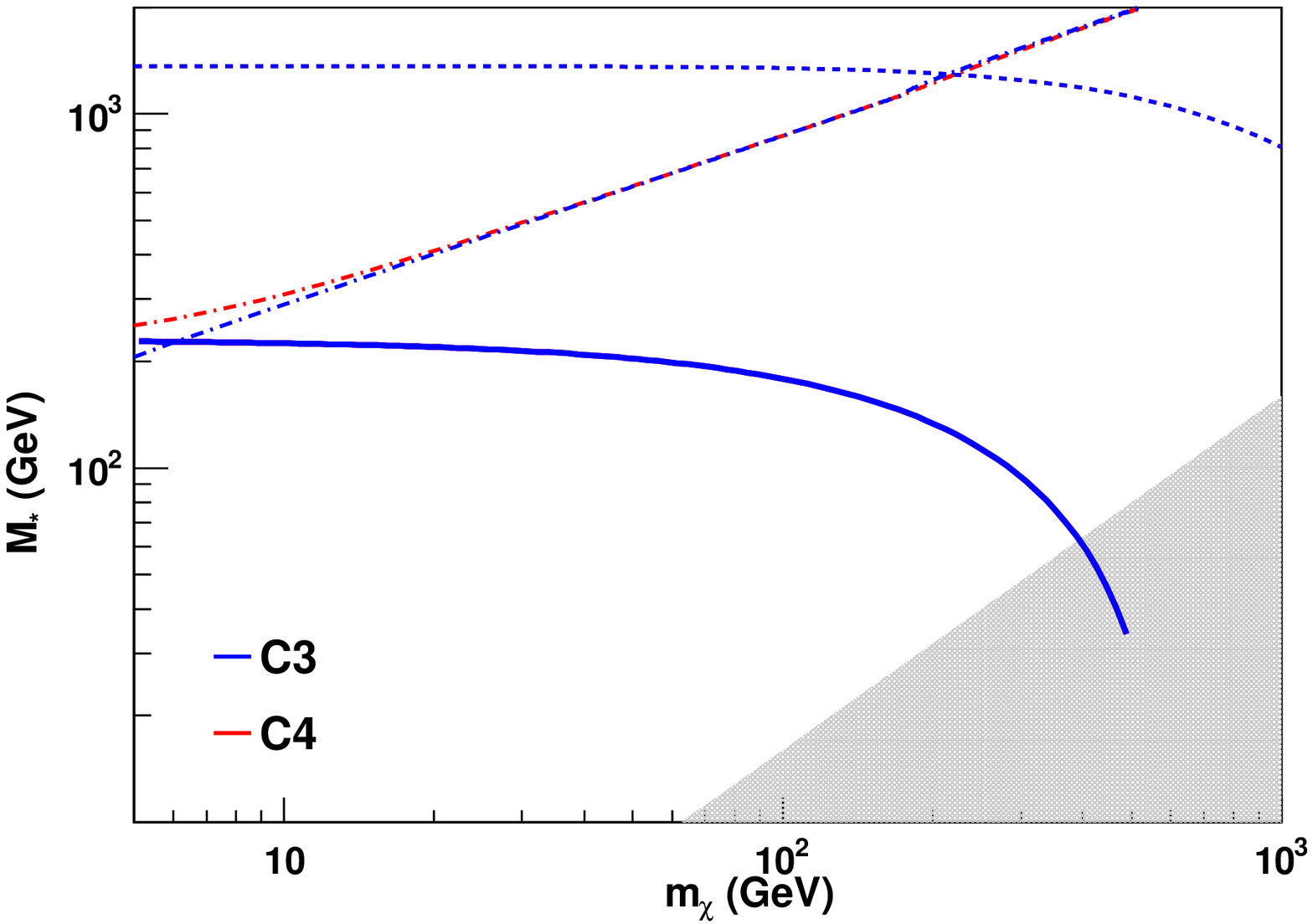}
\caption{\label{fig:C3-4}
Same as Fig.\,\ref{fig:D1-4}, but for the largely degenerate operators C3 and C4.
}
\end{figure}

\begin{figure}[t]
\includegraphics[width=12.0cm]{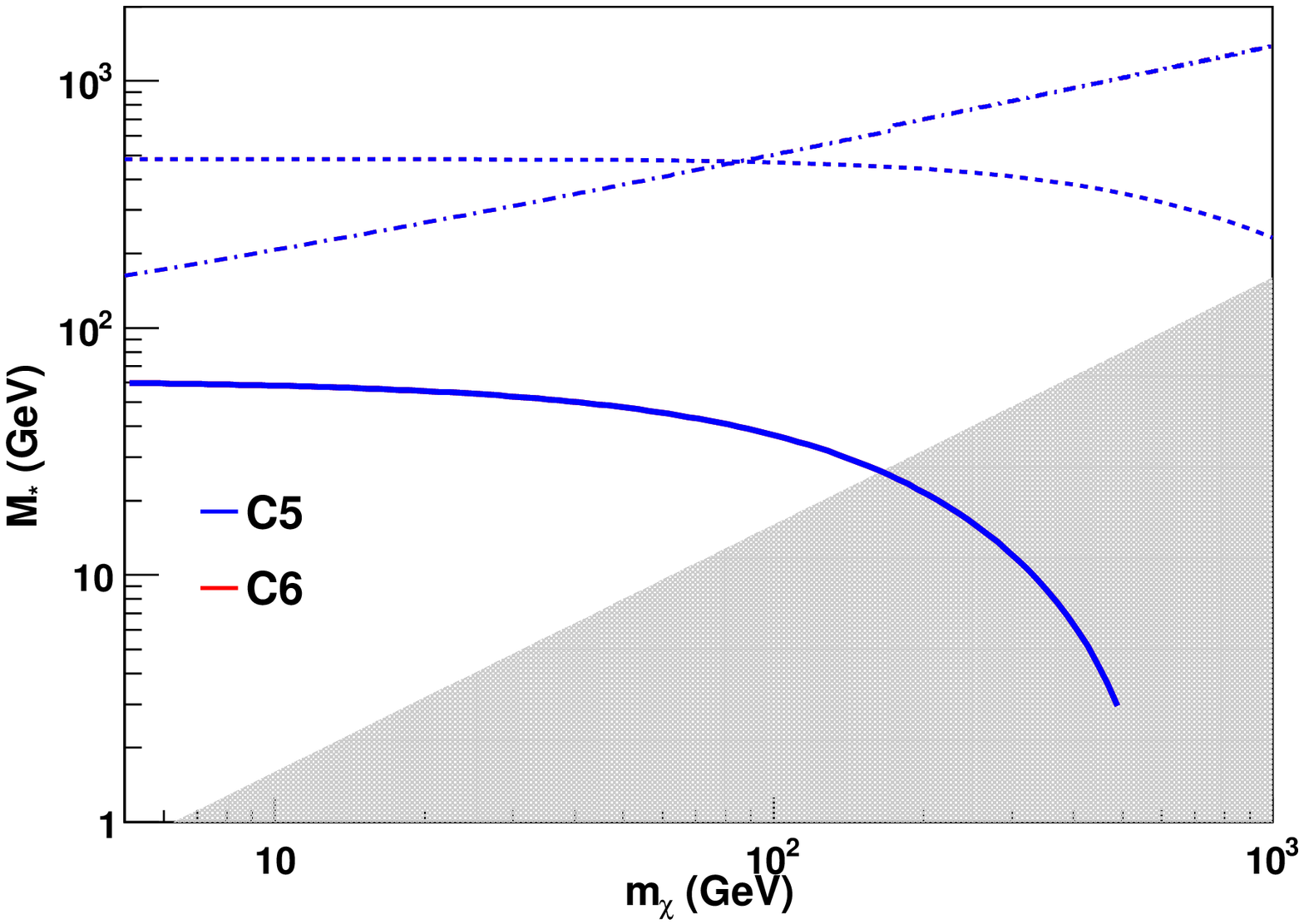}
\caption{\label{fig:C5-6}
Same as Fig.\,\ref{fig:D1-4}, but for the largely degenerate operators C5 and C6.
}
\end{figure}

\begin{figure}[t]
\includegraphics[width=12.0cm]{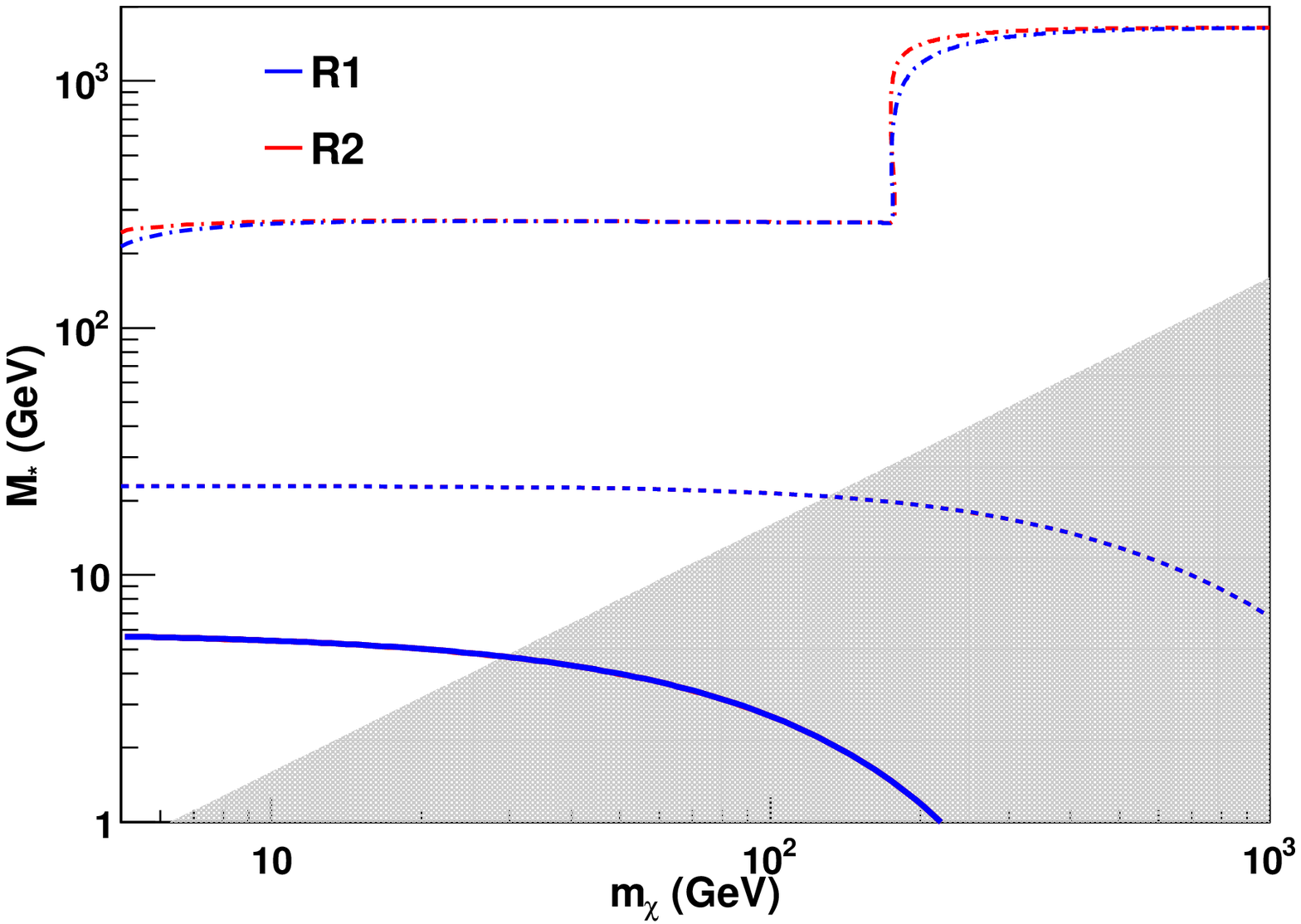}
\caption{\label{fig:R1-2}
Same as Fig.\,\ref{fig:D1-4}, but for the largely degenerate operators R1 and R2.
}
\end{figure}

\begin{figure}[t]
\includegraphics[width=12.0cm]{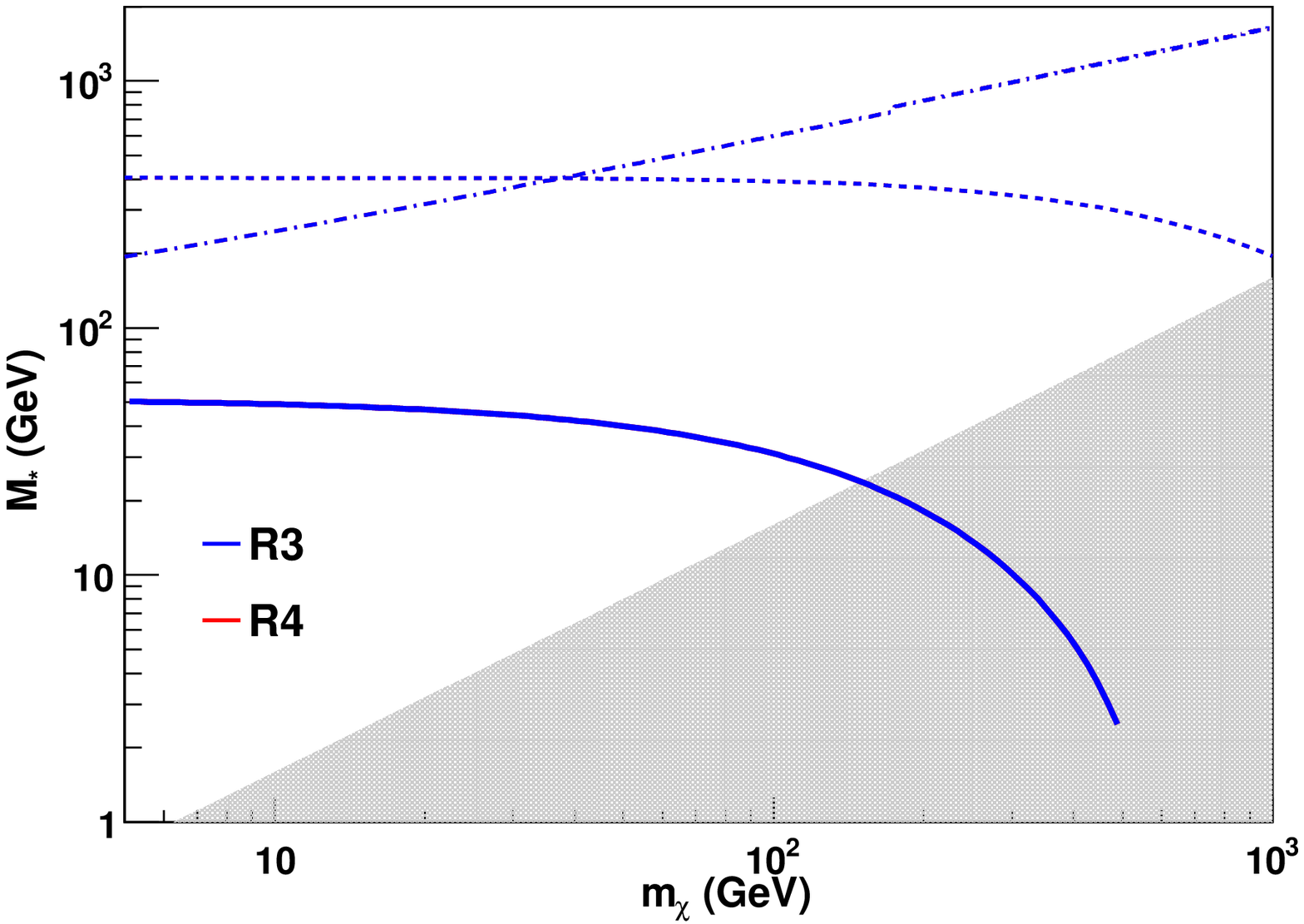}
\caption{\label{fig:R3-4}
Same as Fig.\,\ref{fig:D1-4}, but for the largely degenerate operators R3 and R4.
}
\end{figure}

We plot $2\sigma$ lower limits on the
scale of new physics $M_*$ for each operator as solid lines in
figures~\ref{fig:D1-4}-\ref{fig:R3-4},
where for illustration, we also plot the lines resulting in the
observed thermal relic density.
Comparing with previous studies which analyzed D8 \cite{Beltran:2010ww} and D1, D4, D5, and
D8 (but with a different normalization between different flavors of quarks) \cite{Bai:2010hh},
we find rough agreement with these studies.
It is worth noting that the CDF analysis was somewhat
optimized for theories with large extra dimensions as opposed to $\chi \chi j$,
and it is possible that better bounds may be available for more optimized analysis strategies.

\subsection{LHC Prospects}

We also simulate the inclusive jets and missing transverse energy events at the LHC
experiments for $\sqrt{s} = 14$\,TeV
and compare them with the analysis in Ref.\,\cite{Vacavant:2001sd},
which studied this signal in the context of a large extra dimensions search.
In Ref.\,\cite{Vacavant:2001sd}, the
missing $E_t$ was required to be larger than $500$\,GeV and no veto on additional
hadronic radiation was imposed.  Additional hard isolated leptons were vetoed in order to
reduce backgrounds from $W$ + jets processes.  Finally,
the azimuthal angle between the the missing transverse energy and the second hardest
jet was required to be $\Delta \Phi \geq 0.5$, to mitigate the QCD background due to
mismeasured jet energies.
The expected number of background events after these cuts
was simulated to be about $B=2\times10^4$  with  $100$\,fb$^{-1}$ of data.

Proceeding as before, we find the parton-level cross section $\sigma_{j\chi\chi}$
(with $E_t$ of the jet greater than $500$~GeV) using Comphep.
We use Pythia and the PGS simulation with the generic LHC detector model to estimate
an efficiency, defined as for the Tevatron study, of roughly $80\%$.  This is roughly in
agreement with the efficiency for the monojet signal from the large extra dimension
model, which was found in Ref.~\cite{Vacavant:2001sd} to be $\sim 90\%$.

We define the $5 \sigma$ detection region for which the number
of expected signal events $S$ and background events $B$ satisfy $S / \sqrt{B} \geq 5$
for an integrated luminosity of $100~{\rm fb}^{-1}$.  We
plot this $5\sigma$ reach as dashed lines in figures~\ref{fig:D1-4}-\ref{fig:R3-4}.  Again,
we note that this search was optimized for a large extra dimensions signal, and our
knowledge of the LHC detector performance and SM backgrounds are expected to substantially
improve over those available to Ref.~\cite{Vacavant:2001sd}.  Given our positive results,
a dedicated reanalysis by the collaborations would be very interesting.

\section{Implications for Direct Detection}
\label{sec:direct}

Our new bounds on the strength of interactions of WIMPs with hadrons
can be translated into constraints on the possible contributions to direct
detection cross sections for each of those interactions. Only some operators
contribute to direct detection in the limit of low momentum transfer, and the
remaining operators are suppressed by powers of the WIMP velocity, generically
expected to be
of order $\sim10^{-3}$. For each contributing operator we employ
the expectation value of the partonic operator in the nucleon  \cite{Belanger:2008}.
This, combined with the kinematics of WIMP-nucleon scattering, results in cross sections
\bea
\sigma^{D1}_0&=&1.60\times10^{-37} {\rm cm}^2\left(\frac{\mu_\chi}{1 {\rm GeV}}\right)^2
\left(\frac{20 {\rm GeV}}{M_*}\right)^6,\\
\sigma_0^{D5,C3}&=&1.38\times10^{-37} {\rm cm}^2\left(\frac{\mu_\chi}{1 {\rm GeV}}\right)^2
\left(\frac{300 {\rm GeV}}{M_*}\right)^4,\\
\sigma^{D8,D9}_0&=&9.18\times10^{-40} {\rm cm}^2\left(\frac{\mu_\chi}{1 {\rm GeV}}\right)^2
\left(\frac{300 {\rm GeV}}{M_*}\right)^4,\\
\sigma_0^{D11}&=&3.83\times10^{-41} {\rm cm}^2\left(\frac{\mu_\chi}{1 {\rm GeV}}\right)^2
\left(\frac{100 {\rm GeV}}{M_*}\right)^6,\\
\sigma^{C1,R1}_0&=&2.56\times10^{-36} {\rm cm}^2\left(\frac{\mu_\chi}{1 {\rm GeV}}\right)^2
\left(\frac{10 {\rm GeV}}{m_\chi}\right)^2\left(\frac{10 {\rm GeV}}{M_*}\right)^4,\\
%\sigma_0^{C3}&=&8.72\times10^{-40} {\rm cm}^2\left(\frac{\mu_\chi}{1 {\rm GeV}}\right)^2
%\left(\frac{10 {\rm GeV}}{m_\chi}\right)^2\left(\frac{100 {\rm GeV}}{M_*}\right)^4,\\
\sigma_0^{C5,R3}&=&7.40\times10^{-39} {\rm cm}^2\left(\frac{\mu_\chi}{1 {\rm GeV}}\right)^2
\left(\frac{10 {\rm GeV}}{m_\chi}\right)^2\left(\frac{60 {\rm GeV}}{M_*}\right)^4.
\eea
where $\mu_\chi$ is the reduced mass of the WIMP-nucleon system.

The behavior at low WIMP masses is affected strongly by the spin of the
WIMP itself. For a fermion WIMP the direct detection cross section
for a fixed coupling is largely flat as the
WIMP mass is decreased, until the WIMP is lighter than the nucleon mass.
For a scalar WIMP, except for the vector-type interaction C3, the mass
appears explicitly in the expression for the cross section, causing
the cross section for smaller WIMP masses to notably increase
(provided $m_\chi<M_*$). This has the net effect of weakening the impact
of the collider bounds on the direct detection parameter space for
of very light scalar dark matter with respect to those for
fermion dark matter.

We also notice that the collider bounds on direct detection of non-self-conjugate
fields are stronger by a factor of 2 in cross section than those on
self-conjugate fields. This is an expected result, as the phase space for
direct detection is unchanged by this factor but the phase space of collider
production is suppressed by a factor of 2 for self-conjugate fields.

\begin{figure}[t]
\includegraphics[width=12.0cm]{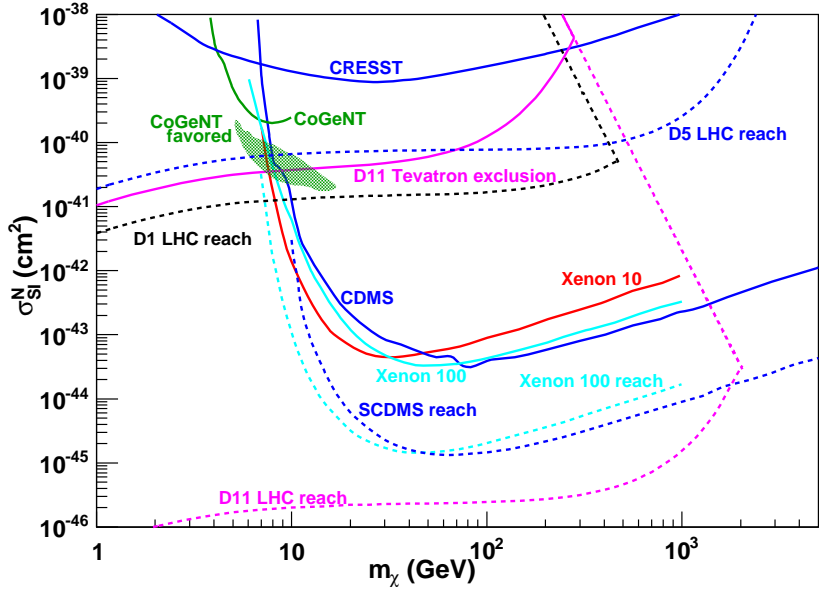}
\caption{\label{fig:DSIplot}
Current experimental limits on spin-independent WIMP direct detection from CRESST \cite{Angloher:2002in}, CDMS \cite{Ahmed:2009zw}, Xenon 10 \cite{Angle:2007uj}, CoGeNT \cite{Aalseth:2010vx}, and Xenon 100 \cite{Aprile:2010um}, (solid lines as labeled), as well as the CoGeNT favored region \cite{Aalseth:2010vx} and future reach estimates for SCDMS \cite{Akerib:2006rr} and Xenon 100 \cite{Aprile:2009yh}, where we have chosen the line using a threshold of 3PE and the conservative extrapolation of $\mathcal{L}_{eff}$ (dashed lines as labeled). Also shown are the current Tevatron exclusion for the operator D11 (solid magenta line) as well as LHC discovery reaches (dashed lines as labeled) for relevant operators.
}
\end{figure}

\begin{figure}[t]
\includegraphics[width=12.0cm]{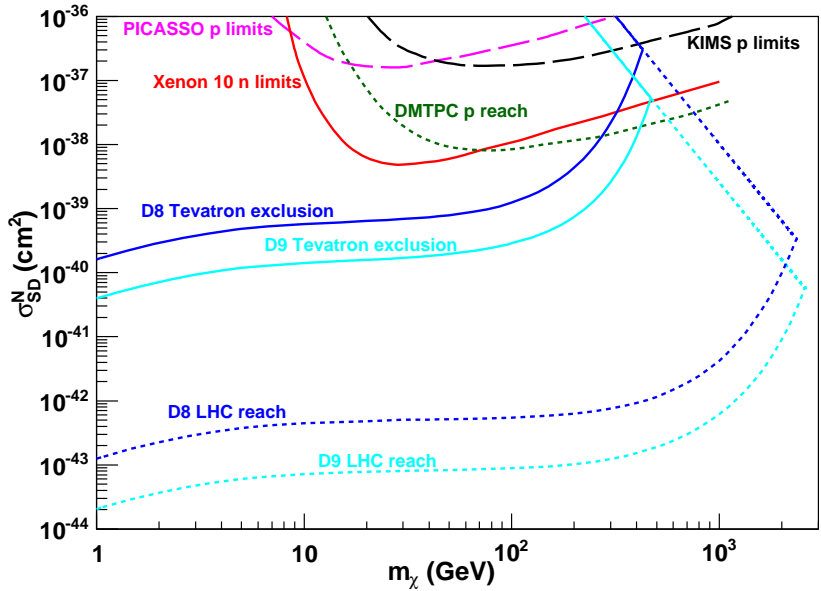}
\caption{\label{fig:DSDplot}
Current experimental limits on spin-dependent WIMP direct detection from Picasso \cite{Archambault:2009sm}, KIMS \cite{Lee.:2007qn}, and Xenon 10 \cite{Angle:2007uj}, as well as the future reach of DMTPC \cite{Sciolla:2009fb}. Also shown are the current Tevatron exclusions (solid lines as labeled) and LHC discovery reaches (dashed lines as labeled) for relevant operators.
}
\end{figure}

\begin{figure}[t]
\includegraphics[width=12.0cm]{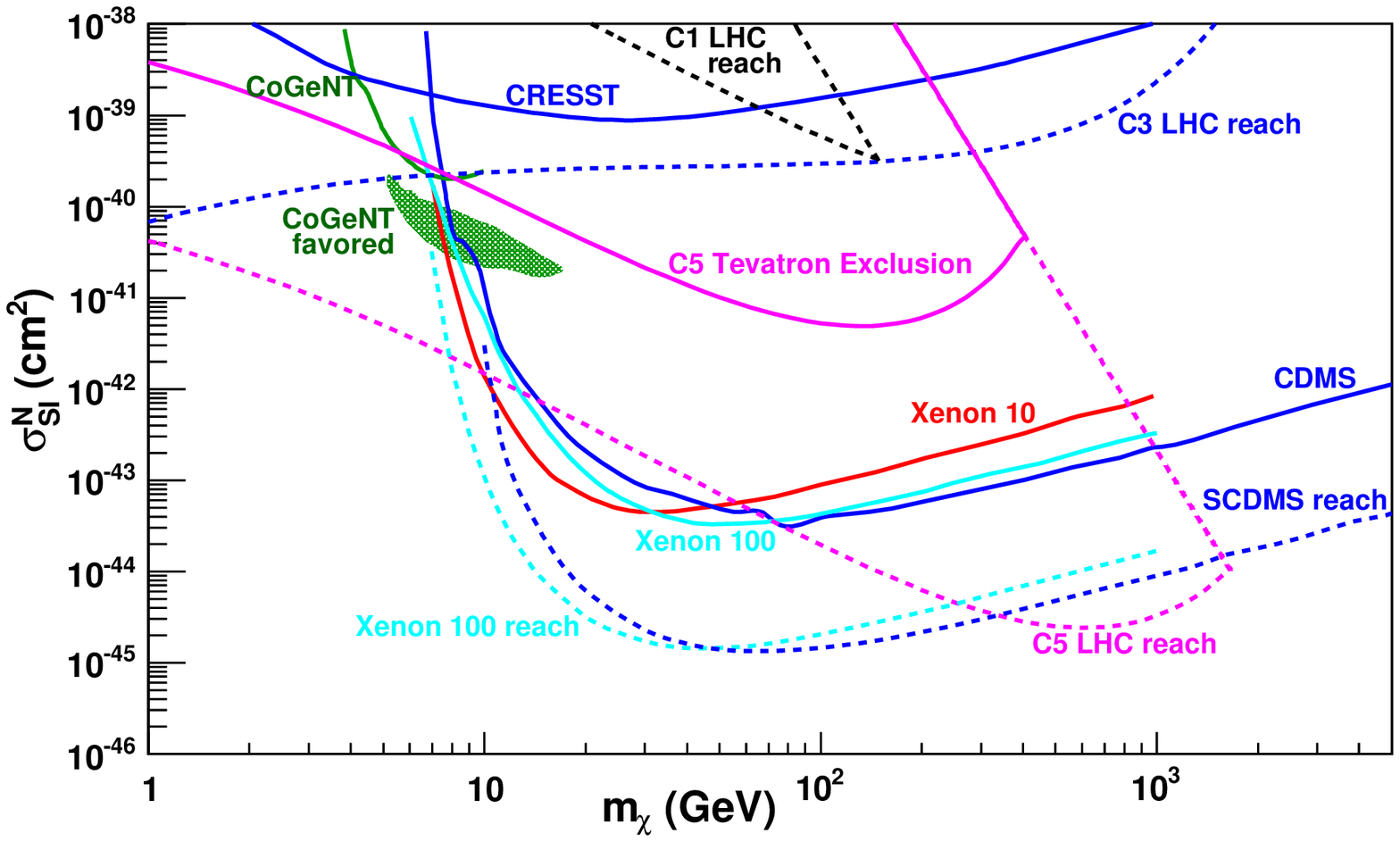}
\caption{\label{fig:CSIplot}
Same as Fig.\,\ref{fig:DSIplot}, but with Tevatron exclusions and LHC reaches for complex scalar WIMP operators, as labeled.
}
\end{figure}

\begin{figure}[t]
\includegraphics[width=12.0cm]{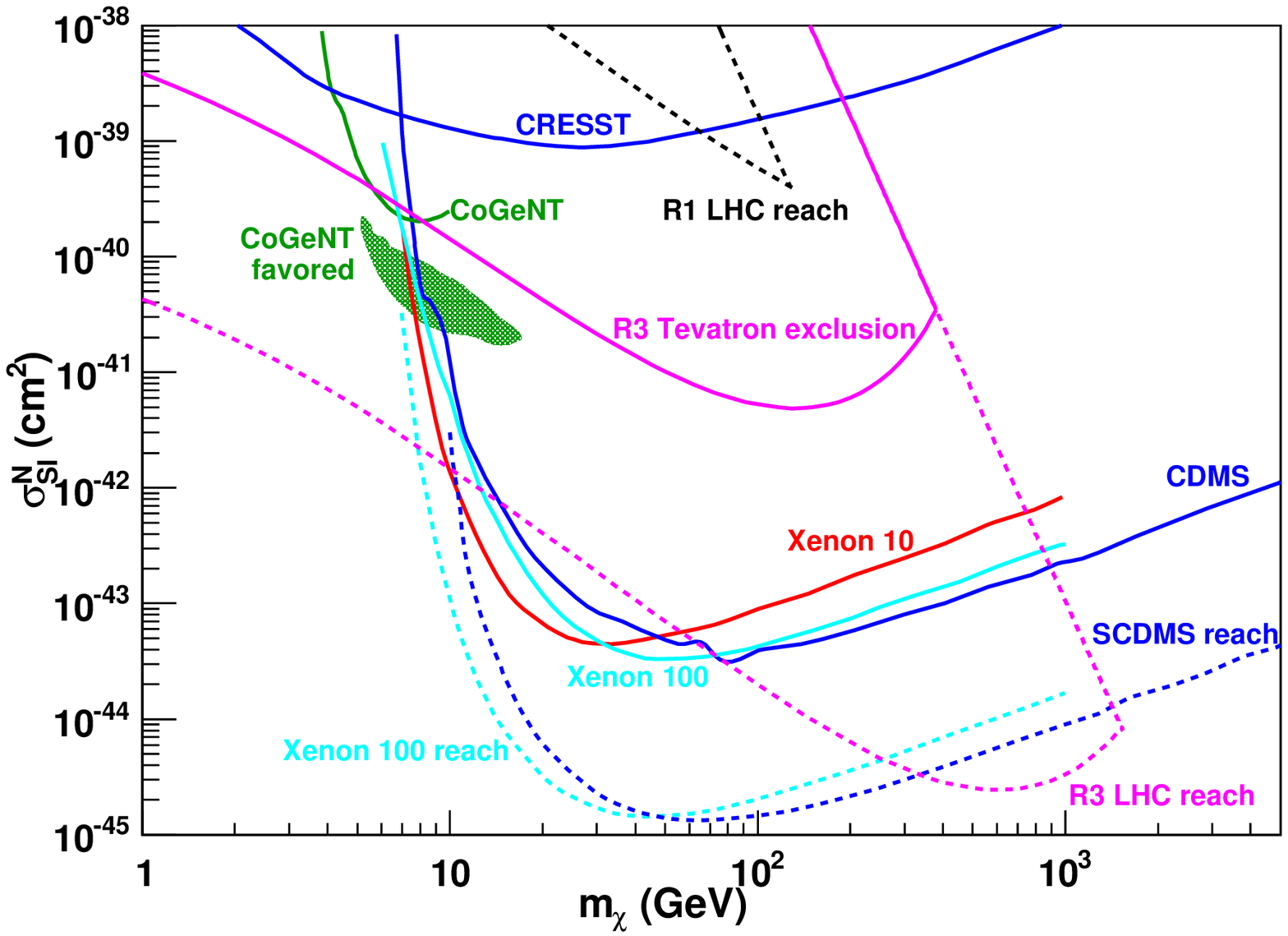}
\caption{\label{fig:RSIplot}
Same as Fig.\,\ref{fig:DSIplot}, but with Tevatron exclusions and LHC reaches for real scalar WIMP operators, as labeled.
}
\end{figure}

\begin{figure}[t]
\includegraphics[width=12.0cm]{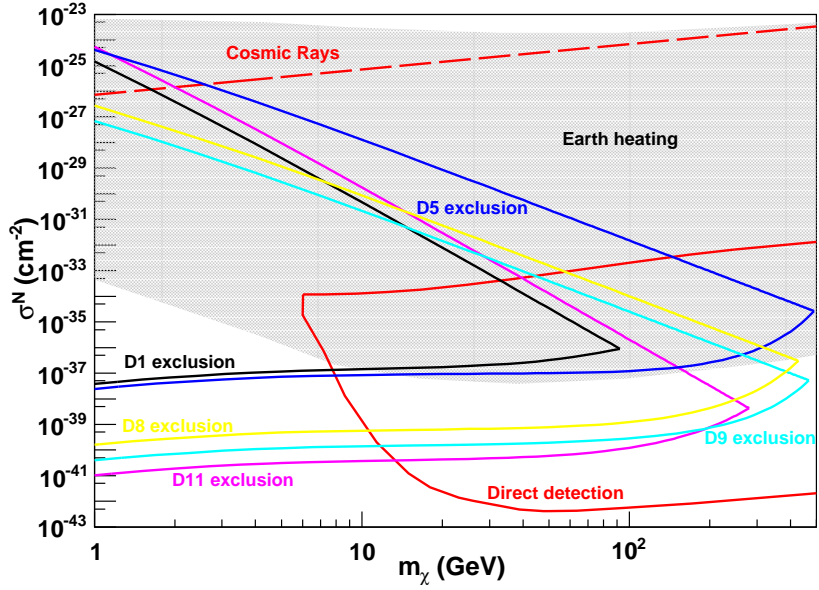}
\caption{\label{fig:Dzoomout}
A cartoon representation of previous limits due to direct detection experiments as well as constraints from Earth heating and cosmic rays\,\cite{Mack:2007xj}, with new exclusions from Tevatron searches for Dirac WIMPs superimposed.
}
\end{figure}

\begin{figure}[t]
\includegraphics[width=12.0cm]{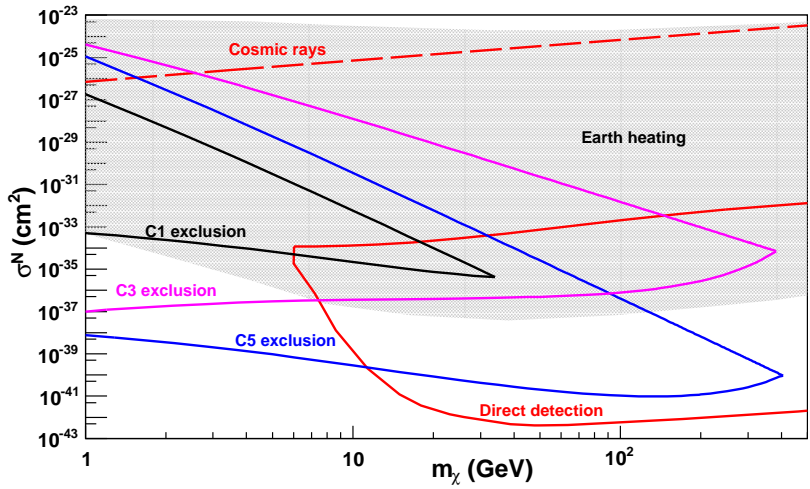}
\caption{\label{fig:Czoomout}
Same as Fig.\,\ref{fig:Dzoomout}, but with complex scalar WIMP limits superimposed.
}
\end{figure}

\begin{figure}[t]
\includegraphics[width=12.0cm]{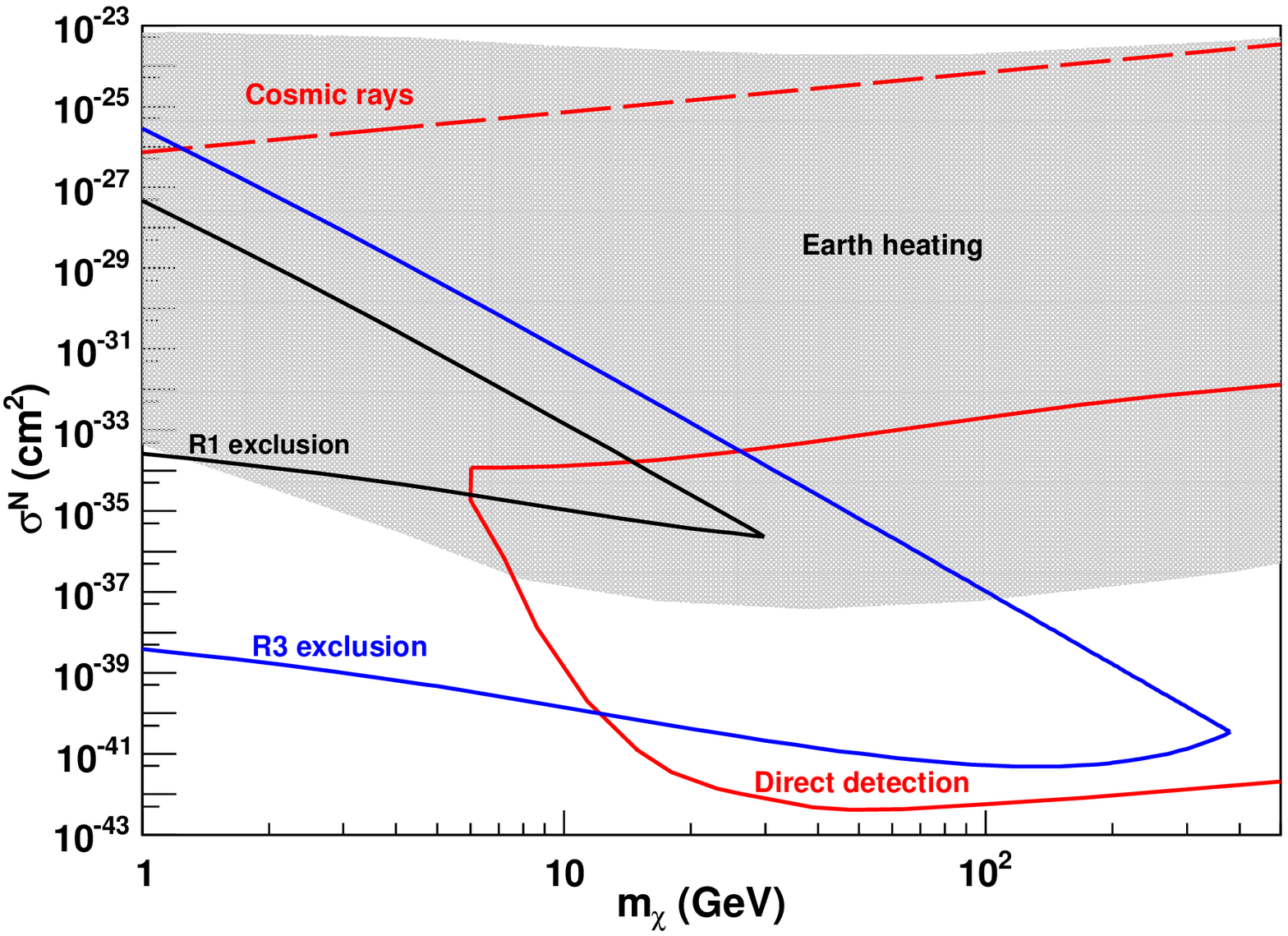}
\caption{\label{fig:Rzoomout}
Same as Fig.\,\ref{fig:Dzoomout}, but with real scalar WIMP limits superimposed.
}
\end{figure}

We plot the effective collider constraints on the WIMP direct detection parameter space in figures
\ref{fig:DSIplot}-\ref{fig:Rzoomout}, including the most relevant direct detection constraints
for comparison.  We see that in all cases, colliders can probe regions of very light WIMP masses
more effectively than direct detection experiments, which become limited by energy thresholds
for extremely light WIMPs.  Indeed, for many operators, the direct detection rates are expected
to be very small because of the velocity suppression, and colliders become the only way
to effectively probe WIMP-hadron interactions.
In the case of a WIMP whose dominant recoil is through a
spin-dependent interaction, collider constraints are already much stronger even than the
expected reaches of near-future direct detection experiments.  Thus, if such an experiment
were to observe a positive signal, the collider constraints would immediately imply a
break-down of the effective field theory at collider energies, revealing the existence of
a light mediator particle.

\section{Conclusions and Outlook}
\label{sec:outlook}

We have studied constraints on dark matter models coming from collider experiments, specifically from extant searches at the Tevatron and future searches at the LHC.
This extends our previous work where we considered the case of dark matter
particles which were Majorana fermions to the cases where the dark matter
is a Dirac fermion or a scalar (either real or complex).

Our results are qualitatively similar to our previous paper. In general collider constraints
are very strong for lighter dark matter and fall off when the dark matter mass exceeds the typical
energy reach of the collider.
The constraints also depend on the coupling of the dark matter; if the dark matter primarily couples
to gluons, the constraints from colliders become especially strong.

One of the most interesting results is that collider constraints on spin dependent interactions are
stronger than direct searches over a
significant portion of parameter space. In the event that direct searches find a signal in this region
while no signal is found at colliders, this will suggest that dark matter is a WIMP of spin 1 or higher,
or that there exists a light mediator particle UV completing the interaction
operators in such a way as to weaken the collider bounds.  The case of a light mediator
with a particular
\begin{center}
dark matter + dark matter  $\leftrightarrow$ SM-neutral mediator $\leftrightarrow$ SM + SM
\end{center}
completion
structure was considered in \cite{Bai:2010hh}.  Beyond these particular constructions,
many models have additional light states which UV complete the
interactions between the dark matter and the Standard Model through a
\begin{center}
dark matter + SM $\leftrightarrow$ SM-charged mediator $\leftrightarrow$ dark matter + SM
\end{center}
topology.
It would be
relatively simple to consider a complete set (as dictated by SM gauge and
Lorentz invariance) of UV completions, and it would be interesting to
see how our bounds are modified in the
presence of such new states, and whether new collider signals can be found to
constrain such models. We leave detailed exploration of these issues for future work.

Finally, we note that while effective theories may not always capture our favorite parameters
of our favorite UV-complete models, they do provide a language to describe
WIMP-SM interactions which captures a wide
class of theories in a fairly model-independent fashion.

 \section*{Acknowledgements}
T. Tait is glad to acknowledge conversations
with M.~Beltran, P.~Fox, D.~Hooper, E. W.~Kolb, Z.~Krusberg, J. Wacker.
and the hospitality of the SLAC theory group, for their generosity during his many visits.
The authors are grateful to E.W.~Kolb and Z.~Kruseberg for catching an error in Eq.~(7) in an
earlier version of the draft.
The work of AR and MI is supported in part by NSF grant PHY-0653656. The work of HY is
supported
in part by NSF grants PHY-0653656 and PHY-0709742.

\end{document}